\documentclass[twocolumn,showpacs,preprintnumbers,amsmath,amssymb]{revtex4}

\usepackage{graphicx}
\usepackage{amsmath} 

\begin{document}

\title{Phase behavior of ditethered nanospheres}

\author{Christopher R. Iacovella$^1$}
\author{Sharon C. Glotzer$^{1,2}$}
\affiliation{$^1$Department of Chemical Engineering and $^2$Department of Materials Science and Engineering \\University of Michigan, Ann Arbor, Michigan 48109-2136}

\date{\today}


\begin{abstract}
We report the results from a computational study of the self-assembly of amphiphilic ditethered nanospheres using molecular simulation. We explore the phase behavior as a function of nanosphere diameter, interaction strength, and directionality of the tether-tether interactions.  We predict the formation of seven distinct ordered phases.   We compare these structures with those observed in linear and star triblock copolymer systems. 
\end{abstract}

\maketitle

\section{Introduction}
The ability to create ordered structures on the nanoscale has important implications for the fabrication of novel electronic \cite{hermanson2001,gyroidsolarcell} and photonic \cite{zhang2006, shevchenko2008} materials.  To this extent, the literature contains a variety of approaches to the assembly process.  For example, dipole moments have been exploited to induce nanoparticles to form sheets \cite{zhang2007, tang2006} and wires \cite{tang2002, murraydipole2005, zhang2007}, binary nanoparticle superlattices reminiscent of atomic crystals have by created by controlling parameters such as particle size ratio and charge ratio \cite{murray2006, shevchenko2008}, and block copolymers have been used as templates to create high conductivity wires \cite{gyroidsolarcell} and other ordered arrays of nanoparticles \cite{knorowski2008}.  The literature also contains many examples of the use of polymer-tethered nanoparticles as a means to self-assemble ordered structures \cite{zhang2003, horsch2005, horsch2006, iacovella2005, iacovella2007, iacovella2008, iacovella2009, arthi2008, arthi2008b, wilson2009, devries2007, devries2008, nie2007, frank2005, song2003, song2003b, trung2008, chan2005, chan2006, zhangx2005, glotzer2005}.  Tethered nanoparticles are hybrid nanoparticle-polymer building blocks where nanoparticles are bonded to immiscible polymer tethers creating amphiphilic building blocks \cite{zhang2003}. The immiscibility between the nanoparticle and polymer tether facilitates microphase separation into bulk periodic structures similar to those observed in block copolymers, but with additional ordering arising from the nanoparticle shape \cite{zhang2003, horsch2005, iacovella2007, trung2008, chan2005,zhangx2005}.

In reference \cite{iacovella2009} we studied systems of ditethered nanospheres (DTNS) where two chemically distinct polymer tethers were attached to the surface of a nanosphere, creating a building block with multiple levels of anisotropy \cite{glotzer2007}.  The DTNS building block is of particular interest as the synthesis of nanospheres functionalized with diametrically opposed tethers has recently been demonstrated, resulting in the formation of wires under dilute conditions \cite{devries2007, devries2008} and preliminary indications of cylinders \cite{stellacci}, as predicted in earlier simulations \cite{zhang2003}.  Just as mono-tethered nanoparticles have been shown to form phases similar to diblock copolymers \cite{horsch2005, iacovella2007, zhang2003} and single-tailed surfactants \cite{iacovella2005, zhang2003}, DTNS have similar analogy with double-tailed surfactants and triblock copolymers \cite{iacovella2009}.  DTNS can be loosely thought of as a nanoparticle equivalent of an ABC triblock copolymer where the center block of a triblock has been replaced by a nanosphere.  In our previous study of DTNS, we found a variety of structures analogous to triblock copolymers, including the alternating double diamond network and alternating tetragonal cylinders \cite{iacovella2009}.  We also found two novel structure not seen in pure triblock copolymer melts, namely NaCl ordered spherical micelles with a complementary simple cubic network of nanospheres and zincblende ordered spherical micelles with a complementary diamond network of nanospheres \cite{iacovella2009}.  In this work we explore the phase behavior of the DTNS system as a function of nanosphere diameter, immiscibility, and directionality of the tether-tether interactions to better understand the structural behavior, expanding on the previous study in reference \cite{iacovella2009}. 

In section \ref{methodology} we discuss the simulation model, method, and analysis methods used in this work.  In section \ref{resultsD} we present the phase behavior as a function of nanoparticle diameter, immiscibility, and directionality of the tether interactions.  In section \ref{discussion} we discuss these results and in section \ref{conclusion} provide concluding remarks.

\begin{figure}[ht]
\centering
\includegraphics[width=2.25in]{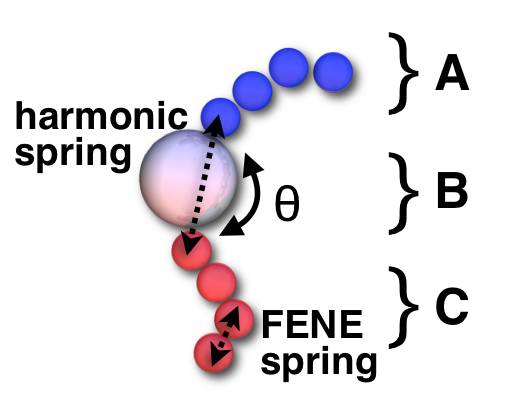} 
\caption{Schematic of the DTNS building block. The blue colored tether is labeled as A, the nanoparticle as B, and the red tether as C. }
\label{DTNS:model}
\end{figure}

\section{Model and Methods }
\label{methodology}
\subsection{Simulation Model and Method}
In the DTNS system, self-assembly is driven by immiscibility and mitigated by (1) the angle between the tethers, (2) the size and shape of the nanoparticle, and (3) the interaction strength between like species.  We utilize a minimal model that captures these important effects and use empirical pair potentials that have been successful in the study of the phase behavior of block copolymers, surfactants, and colloids, and in previous studies of tethered nanoparticles.  

We model polymer tethers as bead spring chains, where individual beads of diameter $\sigma$ are connected via finitely extensible non-linear elastic (FENE) springs  \cite{grest1986} , with the maximum allowable separation set to $R_0$ = 1.5$\sigma$  and the spring constant set to k = 30  \cite{grest1986}. Nanospheres are modeled as beads of diameter $D$ connected to the polymer tethers also via FENE springs.  The planar angle, $\theta$, between tethers at the surface of the nanoparticle is controlled by the use of a harmonic spring, with k = 30 and $R_0 = (D+\sigma)sin(\theta/2)$.  A schematic of the building block is shown in figure \ref{DTNS:model}. We fix the length of the two tethers at 4 beads each, and vary nanosphere diameter, $D$ = 1.5$\sigma$, 2.0$\sigma$, or 2.5$\sigma$ and planar angle, $\theta$ = 30$^\circ$ to 180$^\circ$. 

We use the Brownian dynamics (BD) simulation method \cite{grest1986} to realize long time scales and large systems required to self-assemble complex mesophases. We perform simulations under melt-like conditions where like species attract and unlike species do not attract.  To model the attraction between like species, we use the Lennard-Jones potential (LJ), which induces demixing below a certain critical temperature.  The LJ potential is given by,
 \begin{equation}
U_{LJ} = 
\begin{cases}
4 \epsilon \left( \frac{\sigma^{12}}{(r-\alpha)^{12}}-\frac{\sigma^{6}}{(r-\alpha)^6} \right) + U_{shift}\;, & r-\alpha< r_{cut}\\
0\;, &  r-\alpha \ge r_{cut}
\end{cases}
\label{eqnLJ}
\end{equation}
where $\epsilon$ is the attractive well depth, $U_{shift}$ is the energy at the cutoff, $\alpha$ is the parameter used to shift the interaction to the surface of the nanoparticles to properly account for excluded volume, and $r_{cut}$ is the cutoff of the potential with respect to $\alpha$. $\alpha$ is determined by calculating $\alpha = (D_i - \sigma)/2 + (D_j - \sigma)/2$, where $D_i$ and $D_j$ are the diameters of the two interacting particles.  Note that for interactions between tethers of like species, where $D_i = D_j = \sigma$, $\alpha$ = 0.  The cutoff of the LJ potential is set to $r_{cut} = 2.5\sigma$.  Species of different type interact via a purely repulsive Weeks-Chandler-Andersen (WCA) soft-sphere potential to account for excluded volume interactions.  The WCA potential can be described  by the LJ equation (eqn. \ref{eqnLJ}), with $U_{shift} = \epsilon$ and $r_{cut} = 2^{1/6}\sigma$ \cite{chandler1983}. 

Volume fraction, $\phi$, is defined as the ratio of excluded volume of the beads to the system volume. The degree of immiscibility is determined by the reciprocal temperature, 1/T* =  $\epsilon/k_bT$.  In this study, the potentials are chosen such that they capture the overall nanoparticle-tether and tether-tether immiscibility, the geometry of the nanoparticle, and the angle between the tethers. Changes to the phase behavior are likely if the individual interaction strenghts are changed asymmetrically, but such changes are not within the scope of this study.   

We perform simulations at $\phi = 0.45$ utilizing the following simulation procedure. For a given $\theta$ and $D$ we start from a high-temperature, disordered equilibrated state and incrementally cool the system, allowing the potential energy to equilibrate for several million time steps at each T* before cooling again. For each $\theta$ we perform at least two different cooling sequences in all cases to ascertain the path independence of the observed phase and perform simulations at different system sizes to eliminate artifacts due to box size effects.

\subsection{Method for calculating micelle centers-of-mass}

To analyze structures composed of spherical micelles, it is often necessary to locate the center-of-mass of each micelle.  To approximate the center-of-mass we modify an image processing technique developed by Crocker and Grier that is typically used to identify the center of colloidal particles from microscopy data \cite{crocker1996, varadan2003}.  In this method, we start by constructing a density profile of the system by creating a 3-d grid of cubic cells and calculating the number of particles in each cell. To increase our accuracy, we use approximately 50 snapshots spaced 50 timesteps apart to average over the time-dependent shape of the micelles, but within a small enough time window that the bulk structure is still correlated.   We then multiply the density field by a 3-D Gaussian function (i.e. apply a 3-D Gaussian filter)\cite{crocker1996, varadan2003, cyganek2009} where the diameter of the Gaussian corresponds to the approximate diameter of the micelles.  This weights the center of a spherical object greater than the edges.  We then calculate the cell with the highest weighted value in each region, where a region is defined by the approximate diameter of the micelles. This calculation produces a set of cells that correspond to the centers of mass of the micelles in the system.  To avoid artifacts associated with the resolution of the grid, we vary our cell size between 0.25  and 0.5 .  We additionally calculate the centroid around each center-of-mass cell and then average the values to arrive at the centers used in our analysis.  

\subsection{Mesostructure identification}
To identify the mesophases formed by the DTNS we utilize a combination of visual inspection, calculation of the structure factor, S(q) \cite{schultzthesis2003, structureofmaterials2007}, and construction of the bond order diagram (BOD).  The structure factor is a mathematical description of how a structure scatters incident radiation due the arrangement of material. The structure factor produces a set of strong peaks whose ratio can be used to identify specific geometric structures \cite{structureofmaterials2007}.  The BOD is constructed by taking the directions of all vectors drawn from a particle or micelle to the nearest neighboring particles/micelles.  These vectors are then projected on the surface of a sphere, creating an ``average'' picture of the orientational order in the system.  Systems that have highly correlated nearest neighbor directions (e.g. bulk crystalline materials) will show distinct groups of points on the surface of the sphere. Disordered systems, such as liquids, will appear as points randomly distributed on the surface of the sphere with no clearly correlated points.  For the calculation of these quantities we average over several independent simulation runs and up to 100 time slices.

\section{Phase behavior as a function of $\theta$, $D$ and immiscibility}
\label{resultsD}

\begin{figure}[p]
\begin{center}
\includegraphics[width=3.25in]{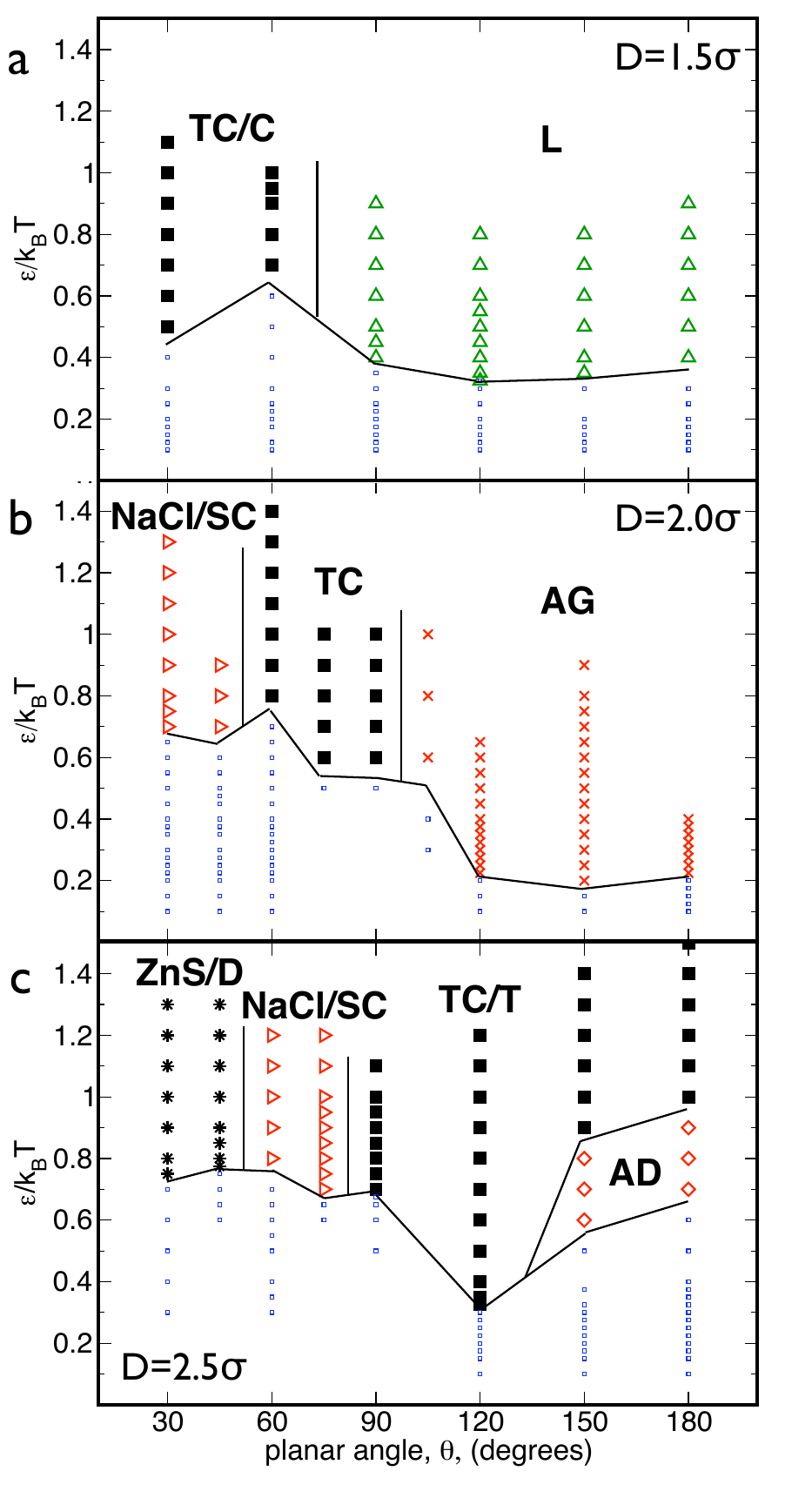} 
\caption{
Phase diagrams as a function of planar angle, $\theta$, and 1/T*, grouped by $D$ .  The structure labels are as follows:  alternating tetragonally ordered cylinders formed by the tethers where the nanospheres also form cylinders (TC/C); alternating tetragonally ordered cylinders formed by the tethers where the nanospheres form a tetragonal mesh (TC/T);  alternating tetragonally ordered cylinders formed by the tethers where the nanospheres form cylinders or a tetragonal mesh (TC); alternating double gyroid structure within a nanosphere matrix (AG); alternating double diamond structure within a nanosphere matrix (AD); lamellar sheets where nanospheres form a monolayer (L/M); NaCl ordered spherical micelles where nanospheres form a complementary simple cubic network (NaCl/SC); and ZnS (i.e. binary diamond) ordered spherical micelles where nanospheres form a complementary diamond network (ZnS/D).  Phase boundaries are approximate and drawn to guide the eye. \textbf{c} is adapted from reference \cite{iacovella2009}.}
\label{DTNS:angle_phasediagram}
\end{center}
\end{figure}

In this section we explore the phase behavior of ditethered nanospheres under melt-like conditions as a function of planar angle, immiscibility, and nanoparticle diameter.  Note that the length of each tether is fixed at 4 beads.   The overall phase behavior, grouped by nanosphere diameter $D$, is summarized in the ``phase diagrams''  shown in figures \ref{DTNS:angle_phasediagram}a-c; over 500 individual statepoints are included in these diagrams.  The individual structures found in figures \ref{DTNS:angle_phasediagram}a-c are discussed in detail in the subsections that follow.   Throughout this chapter, figures are color coded as shown in figure \ref{DTNS:model}; A tethers and the aggregates they form are colored blue, B nanoparticles and the aggregates they form are white/gray, and C tethers and the aggregates they form are red.   

\subsection{Lamellar sheets/nanosphere monolayer (L/M)}
\begin{figure}[h]
\begin{center}
\includegraphics[width=2.25in]{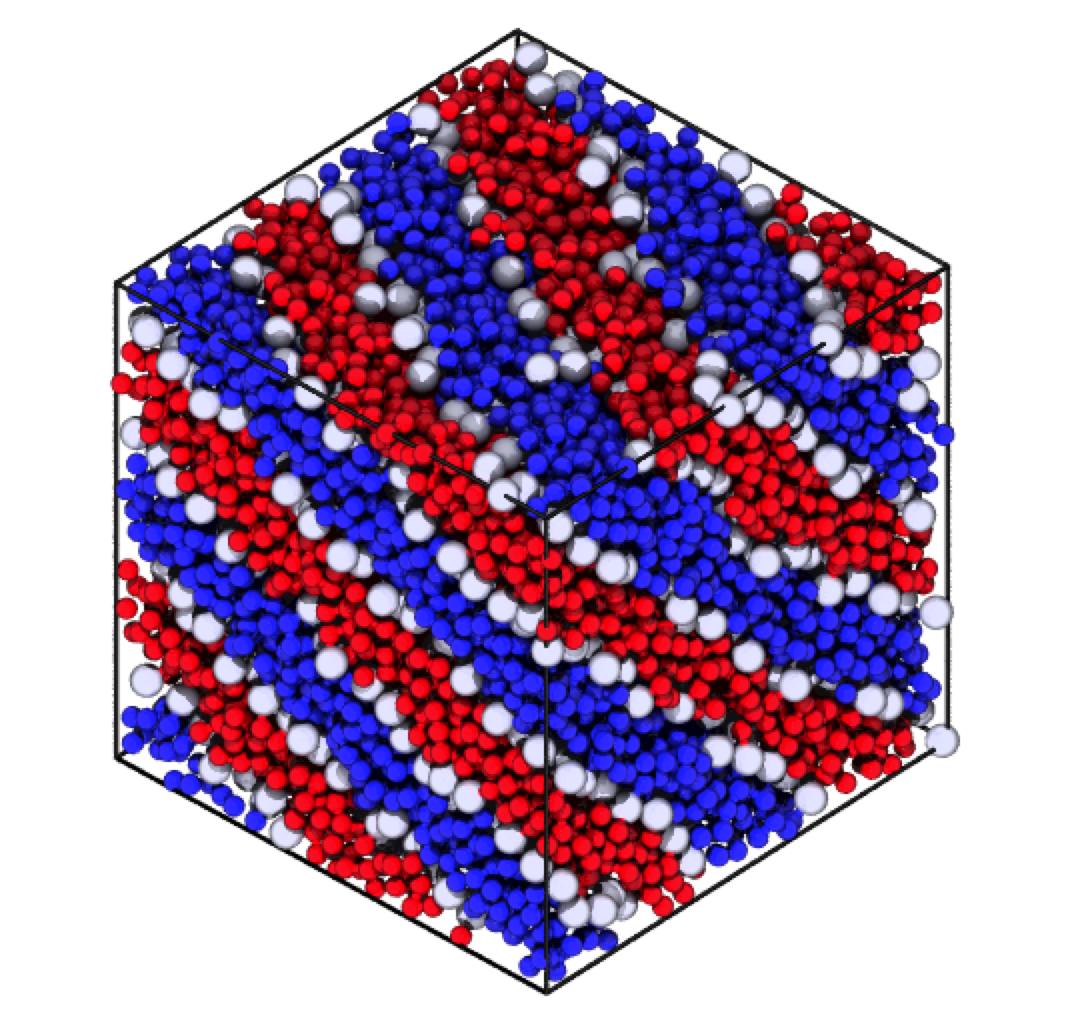} 
\caption{Lamella structure formed for D=1.5$\sigma$, $\theta$=150$^\circ$ and 1/T*= 0.8 showing a monolayer of nanospheres.  The spacing between the nanosphere layers is $\sim$4.2$\sigma$. }
\label{DTNS:d15_lamella}
\end{center}
\end{figure}

A simulation snapshot of a lamellar structure (L/M) for $D$=1.5$\sigma$, $\theta$=150$^\circ$ and 1/T*= 0.8  is shown in figure \ref{DTNS:d15_lamella}.  Nanospheres form monolayers sandwiched between alternating layers of tethers.  The spacing between the layers varies slightly with $\theta$ with the largest spacing occurring for systems where $\theta$=180$^\circ$.  For lamellae, the spacing between layers (i.e. periodicity) can be calculated using $L_{spacing} = (2\Pi/q*)$, where $q*$ is the modulus of the wave vector at which the first principal maximum in S(q) is located \cite{escobedo2005, hajduk1994}.  We find that for $D$=1.5$\sigma$ and $\theta$ = 90$^\circ$, 150$^\circ$, and 180$^\circ$, the average spacing is $L_{spacing} \approx$ 3.9$\sigma$, 4.2$\sigma$, and 4.3$\sigma$, respectively; the spacing does not appear to depend on 1/T* given we are in the ordered regime. 

\subsection{Alternating tetragonally ordered cylinders/nanosphere cylinders (TC/C) or tetragonal mesh (TC/T)}
\begin{figure}[h]
\begin{center}
\includegraphics[width=3.25in]{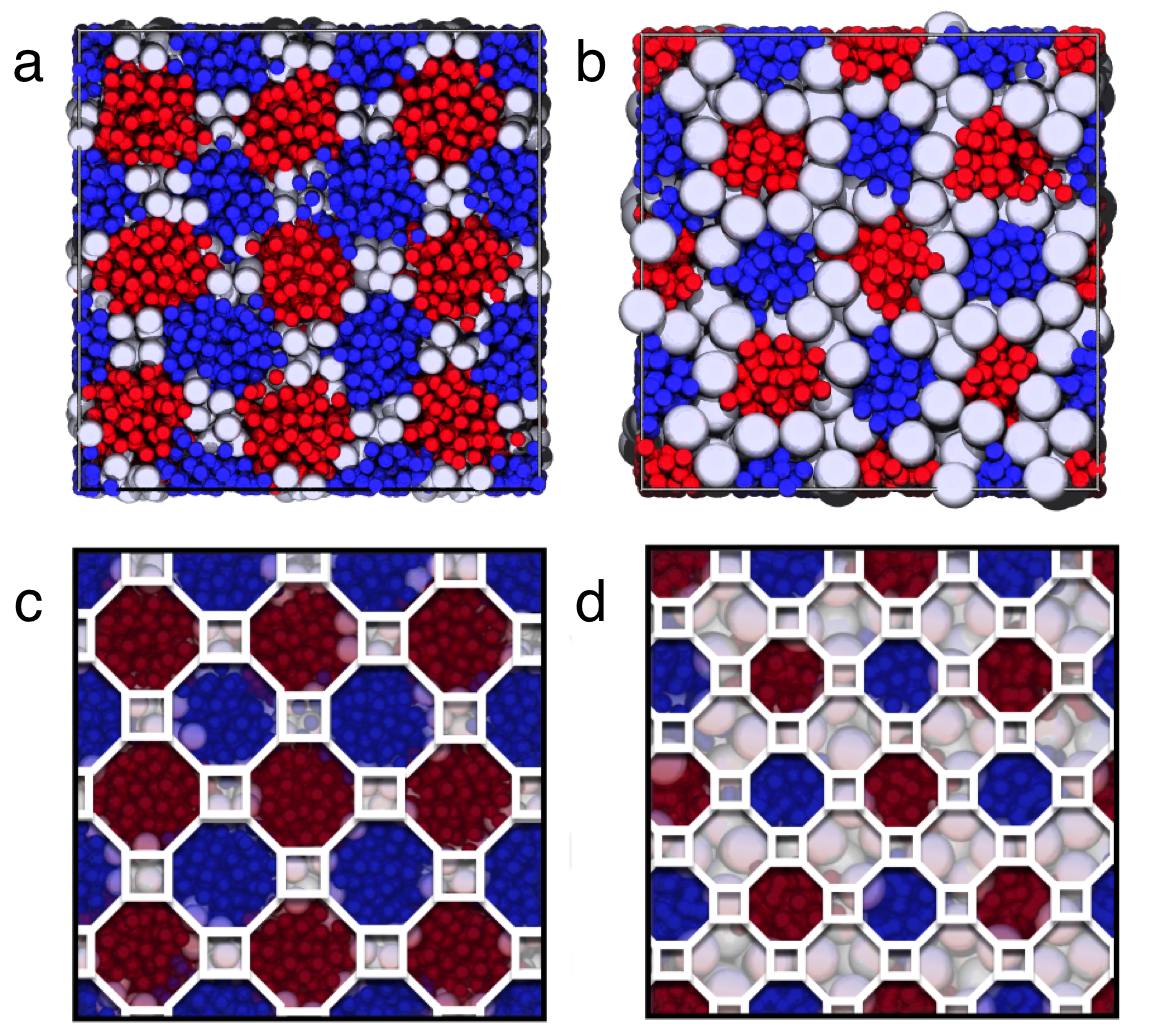} 
\caption{\textbf{(a)} Simulation snapshot of the end view of the TC/C phase for $D$=1.5$\sigma$, $\theta$ = 60$^\circ$, and 1/T* = 0.8.  \textbf{(b)} Simulation snapshot of the end view of the TC/T phase for $D$=2.5$\sigma$, $\theta$ = 90$^\circ$, and 1/T* = 0.8. \textbf{(c)} [8,8,4] Archimedean tiling overlayed on the simulation snapshot in (a).  \textbf{(d)} [8,8,4] Archimedean tiling overlayed on the simulation snapshot in (b).}
\label{DTNS:cylinders}
\end{center}
\end{figure}

Each nanosphere diameter investigated demonstrates a region of cylindrical structures.  In all cases, we find a cross-sectional checkerboard pattern of alternating tetragonally ordered cylinders formed by the tethers.  However, the arrangement of the nanospheres is different when we consider different nanosphere diameters. For $D$=1.5$\sigma$ nanospheres organize into cylinders (TC/C), as shown in figure \ref{DTNS:cylinders}a.  In contrast, for $D$=2.5$\sigma$ nanospheres organize into a tetragonal mesh that separates the alternating cylinders (TC/T), as shown in figure \ref{DTNS:cylinders}b. In both cases, the structures can be described by the [8,8,4] Archimedean tiling constructed of octagons and squares, shown in figure \ref{DTNS:cylinders}c for $D$=1.5$\sigma$ and figure \ref{DTNS:cylinders}d for $D$=2.5$\sigma$.  We should note that for $D$=1.5$\sigma$ tethers occupy the octagonal tiles and nanospheres occupy the square tiles, whereas for $D$=2.5$\sigma$ both tethers and nanospheres occupy the octagonal tiles and only nanospheres occupy the square tiles.  

For $D$=2.0$\sigma$ we see a both tetragonal and cylindrical morphologies of nanospheres.  As the value of 1/T* increases (i.e. we cool the system), there is a gradual transition from a tetragonal mesh of nanospheres to distinct cylinders formed by the nanospheres.  Conceptually, the tetragonal phase can be thought of as the cylinder phase with connections between the cylinders.  As the strength of the attraction between the nanospheres increases (i.e. we cool the system), these connections are minimized forming more compact, cylindrical nanosphere domains.

\subsection{Alternating gyroid/nanosphere matrix (AG)}
\begin{figure}[h]
\begin{center}
\includegraphics[width=3.25in]{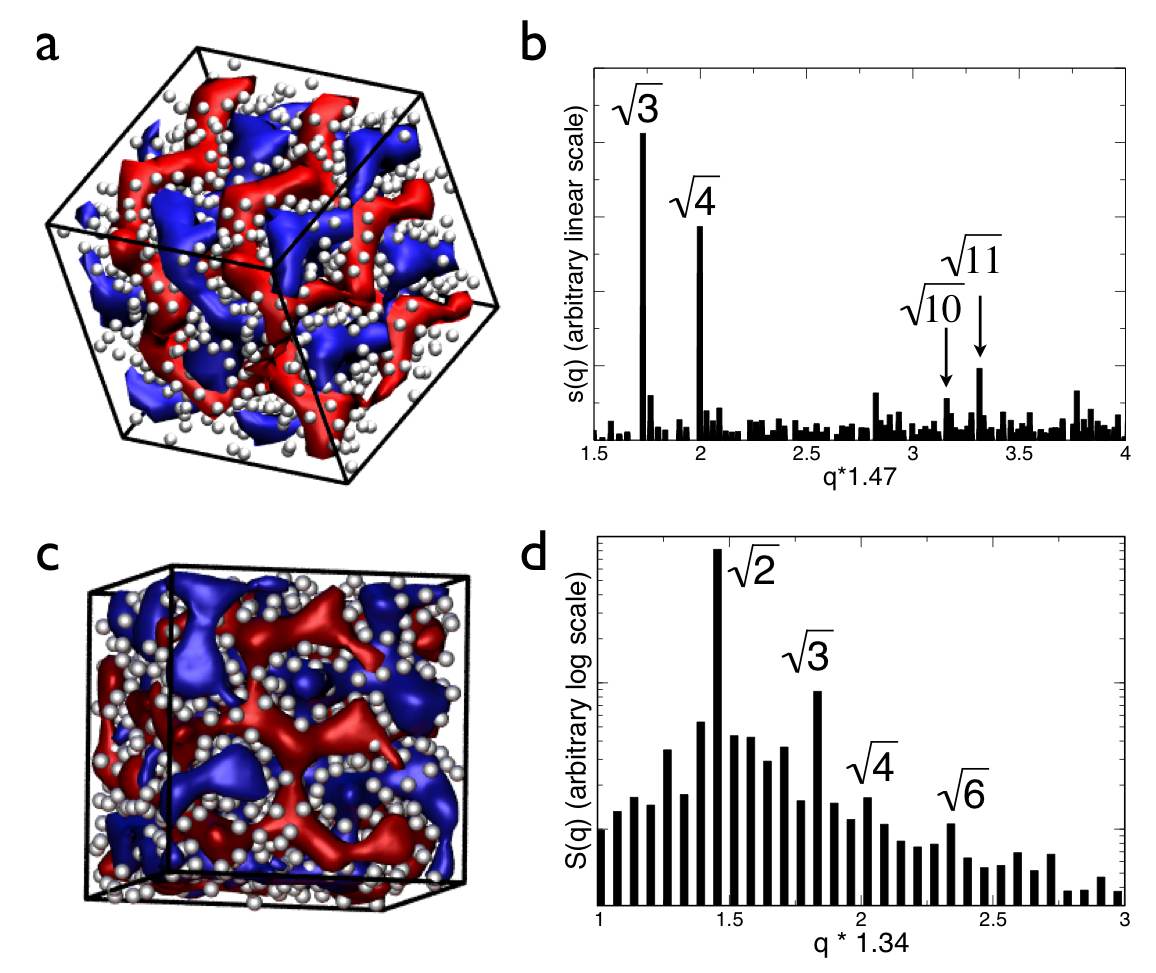} 
\caption{\textbf{(a)} Isosurfaces of the two tether domains, which self-assembled into the AG structure showing 8 unit cells with a unit cell size of ~13$\sigma$; nanospheres are rendered at half their true size. \textbf{(b)} Structure factor of the AG phase.  Data in (a-b) correspond to $D$=2.0$\sigma$, $\theta$  = 180$^\circ$ and 1/T* = 0.4. \textbf{(c)} Isosurfaces of the two tether domains, which self-assembled into the AD structure showing 27 unit cells with a unit cell size of ~10$\sigma$; nanospheres are rendered at half their true size. \textbf{(d)} Structure factor of the AD phase.  Data in (c-d) correspond to D=2.5$\sigma$, $\theta$  = 180$^\circ$ and 1/T* = 0.8 and are adapted from reference \cite{iacovella2009}.}
\label{DTNS:AG_AD}
\end{center}
\end{figure}

A simulation snapshot of the alternating gyroid structure (AG) is shown for $D$=2.0$\sigma$, $\theta$=180$^\circ$, and 1/T* = 0.4 in figure \ref{DTNS:AG_AD}a.  The AG structure consists of two chemically distinct gyroid networks, one network formed by the A tethers and one formed by the C tethers, separated by a matrix of nanospheres.  These networks are composed of a series of interconnected tubes, where three tubes meet at each node.  The AG structure has the same Ia3d space group of the double gyroid (DG) \cite{matsen1996, iacovella2008}, however each gyroid network is chemically distinct in the AG phase.  This structure was identified both visually and by the use of the structure factor, S(q), finding characteristic peaks in ratio $\sqrt{3}:\sqrt{4}:\sqrt{10}:\sqrt{11}$ as expected \cite{escobedo2005} and shown in figure \ref{DTNS:AG_AD}b; for ease of viewing the x-axis was scaled by 1.47  such that the numerical values correspond to the values in the characteristic ratio (i.e. the first peak occurs at a numerical value of  $\sqrt{3}$).

\subsection{Alternating diamond/nanosphere matrix (AD)}
The alternating diamond phase (AD), shown in figure \ref{DTNS:AG_AD}a for $D$=2.5$\sigma$, $\theta$ =180$^\circ$ and 1/T*=0.8, consists of two chemically distinct, interpenetrating diamond networks, one formed by the A tethers and one formed by the C tethers, separated by a matrix of nanospheres. This phase could also be classified as the double diamond structure if we ignore the chemical specificity of the two diamond networks \cite{escobedo2007}. Each diamond network is composed of cylindrical tubes, where four tubes connect at a node in a tetrahedral arrangement.  This structure was identified visually and by calculating S(q).  S(q) for the alternating diamond phase is plotted in figure \ref{DTNS:AG_AD}d, showing characteristic peaks with ratio $\sqrt{2}:\sqrt{3}:\sqrt{4}:\sqrt{6}$  as expected \cite{escobedo2007}; for ease of viewing the x-axis was scaled by 1.34, such that the numerical values correspond to the values in the characteristic ratio (i.e. the first peak occurs at a numerical value of  $\sqrt{2}$).

\subsection{NaCl ordered spherical micelles/nanosphere simple cubic network (NaCl/SC)}
\begin{figure}[h]
\begin{center}
\includegraphics[width=3.25in]{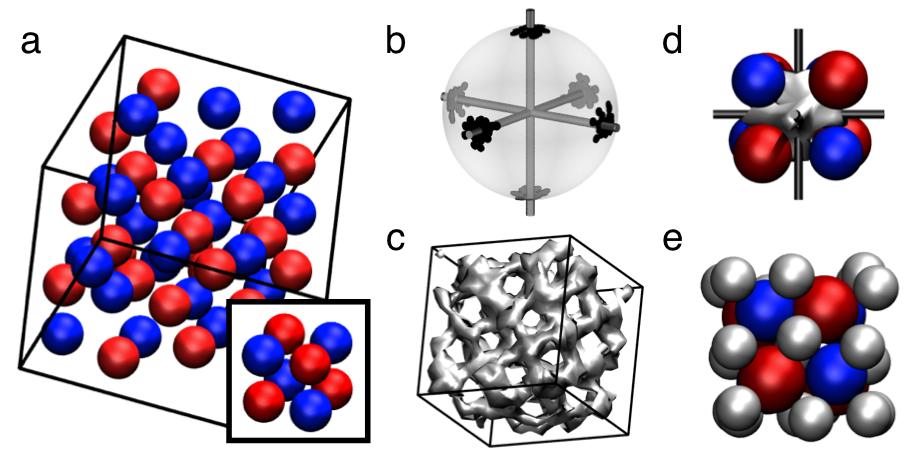} 
\caption{\textbf{(a)} Centers of mass of the NaCl ordered micelles formed by the tethers; the unit cell size is approximately 6.5$\sigma$.  A perfect NaCl unit cell is inset. \textbf{(b)} BOD of the micelles' center of mass; the BOD of a perfect NaCl structure is shown as lines. \textbf{(c)} Isosurface of the nanoparticles showing a simple cubic network arrangement.  \textbf{(d)} Eight particle unit cell of NaCl formed by the micelles in the system, with the node of the nanoparticles network at the interstitial shown as an isosurface. \textbf{(e)} 8 particle unit cell of NaCl formed by the micelles in the system with the nodes of the nanoparticle network drawn as white spheres showing the AlCu$_2$Mn structure.  All data for $D$=2.5$\sigma$, $\theta=60^\circ$ and 1/T*=0.8. Adapted from reference \cite{iacovella2009}.}
\label{DTNS:nacl}
\end{center}
\end{figure}

For $D$=2.0$\sigma$ and 2.5$\sigma$ the tethers form spherical micelles that order into an NaCl lattice with a complementary simple cubic network formed by nanoparticles (NaCl/SC). In figure \ref{DTNS:nacl}a we plot a simulation snapshot of the centers of mass of the micelles formed by the tethers that order into a NaCl lattice for $D$=2.5$\sigma$, $\theta$=60$^\circ$, and 1/T*=0.8. We can see that the structure clearly demonstrates alternating chemical specificity.  Figure \ref{DTNS:nacl}b shows the bond order diagram (BOD) for the centers of mass of the micelles, where we ignore chemical specificity of the micelles and simply calculate the BOD for nearest neighbor micelles. The resulting BOD shows a simple cubic arrangement of the micelles, which corresponds to the BOD of a perfect NaCl system, shown as lines in figure \ref{DTNS:nacl}b. 

The nanoparticles fill the space between the micelles arranging into a simple cubic network; an isosurface of the nanoparticles is shown in figure \ref{DTNS:nacl}d.  The nodes of the simple cubic structure each have six connections points; a single node is shown in figure \ref{DTNS:nacl}e. The highest density of nanoparticles (i.e. the node of the SC network) resides in the center of an eight-particle NaCl unit cell, as shown in figure \ref{DTNS:nacl}e -- this high density location corresponds to the placement of the central particle in a BCC lattice. We approximate the location of the nodes using the micelle  center of mass calculation; figure \ref{DTNS:nacl}f shows an eight-particle unit cell of NaCl extracted from our system with the locations of the nanoparticle nodes rendered as spheres.  The overall phase corresponds to the AlCu$_2$Mn structure, also known as the Heusler (L$_2$1) phase \cite{navysite}.  The AlCu$_2$Mn structure is a three-component analog of BCC (note, CsCl is a two-component analog to BCC).

\subsection{Zincblende ordered spherical micelles/nanosphere diamond network (ZnS/D)}
\begin{figure}[ht]
\begin{center}
\includegraphics[width=3.25in]{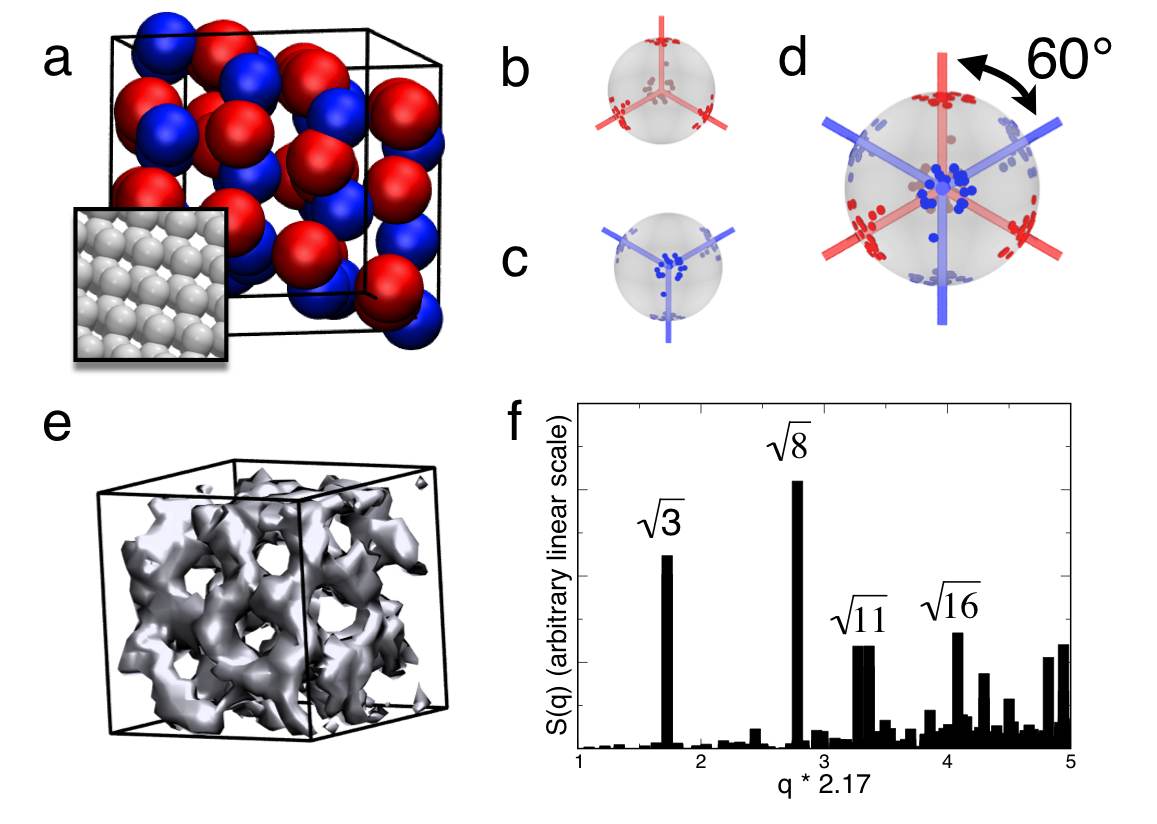} 
\caption{\textbf{(a)} Centers of mass of the micelles formed by tethers that order into the ZnS lattice showing 27 unit cells with a unit cell size of approximately 8$\sigma$.  A perfect diamond lattice is shown in the inset.  \textbf{(b-c)} BOD for nearest neighbors in the ZnS structure for the two different tetrahedral arrangements.  \textbf{(d)} The combination of figures (b) and (c). \textbf{(e)} Diamond network formed by the nanoparticles in the Zns/D phase. \textbf{(f)} Diamond network formed by the nanoparticles in the Zns/D phase. \textbf{(f)} Average S(q) for the nanoparticle network, showing strong peaks with ratios $\sqrt{3}:\sqrt{8}:\sqrt{11}:\sqrt{16}$.  Note that q was scaled such that the first peak corresponds to $\sqrt{3}$ for ease of viewing.  All DTNS data at $D$=2.5$\sigma$, $\theta$  = 30$^\circ$ and 1/T* = 0.8. Adapted from reference \cite{iacovella2009}.}
\label{DTNS:zns}
\end{center}
\end{figure}

For $D$=2.5$\sigma$ we find a binary mixture of spherical micelles formed by the tethers that order into a zincblende (binary diamond) lattice with a complementary diamond network formed by the nanospheres (ZnS/D).  In figure \ref{DTNS:zns}a, we show a simulation snapshot of the centers of mass of the micelles formed by the tethers in the zincblende structure.  This structure is the two-component analog of the diamond lattice \cite{navysite}. Figures \ref{DTNS:zns}b-d show the BODs of the micelles' centers of mass for the zincblende structure.  We split the BOD into two separate diagrams, since the diamond phase possesses two bond configurations that are 60$^\circ$ rotations of each other.  We plot clusters where a ``blue'' particle (i.e. micelle formed by the A portion) is at the center surrounded by ``red'' particles (i.e. micelles formed by the C portion), and a second diagram where clusters have a ``red'' particle at the center surrounded by ``blue'' particles; these cluster definitions properly group the data by orientation of the tetrahedrons.  Both BODs in figure \ref{DTNS:zns}b and c show clear tetrahedral arrangements; we plot the BODs for an ideal zincblende structure as lines in both plots, showing good agreement with our simulation results.  The blue centered and red centered tetrahedrons have complementary orientations (i.e. rotations of 60$^\circ$), as shown in figure \ref{DTNS:zns}d.   

Within this phase, the nanoparticles organize into a diamond network that is woven into the micellar lattice. An isosurface of the nanoparticle diamond network structure is shown in figure \ref{DTNS:zns}e.  In figure \ref{DTNS:zns}f we plot S(q) for the nanoparticles, finding strong peaks at ratios $\sqrt{3}:\sqrt{8}:\sqrt{11}:\sqrt{16}$ as expected for diamond \cite{structureofmaterials2007}; note that q was scaled such that the first peak corresponds to $\sqrt{3}$ for ease of viewing.  The overall ZnS/D phase is composed of two interwoven diamond structures where one network is formed by nanoparticles and the other consists of a binary lattice of spherical micelles. 

\section{Discussion}
\label{discussion}
\begin{figure}[ht]
\begin{center}
\includegraphics[width=3.25in]{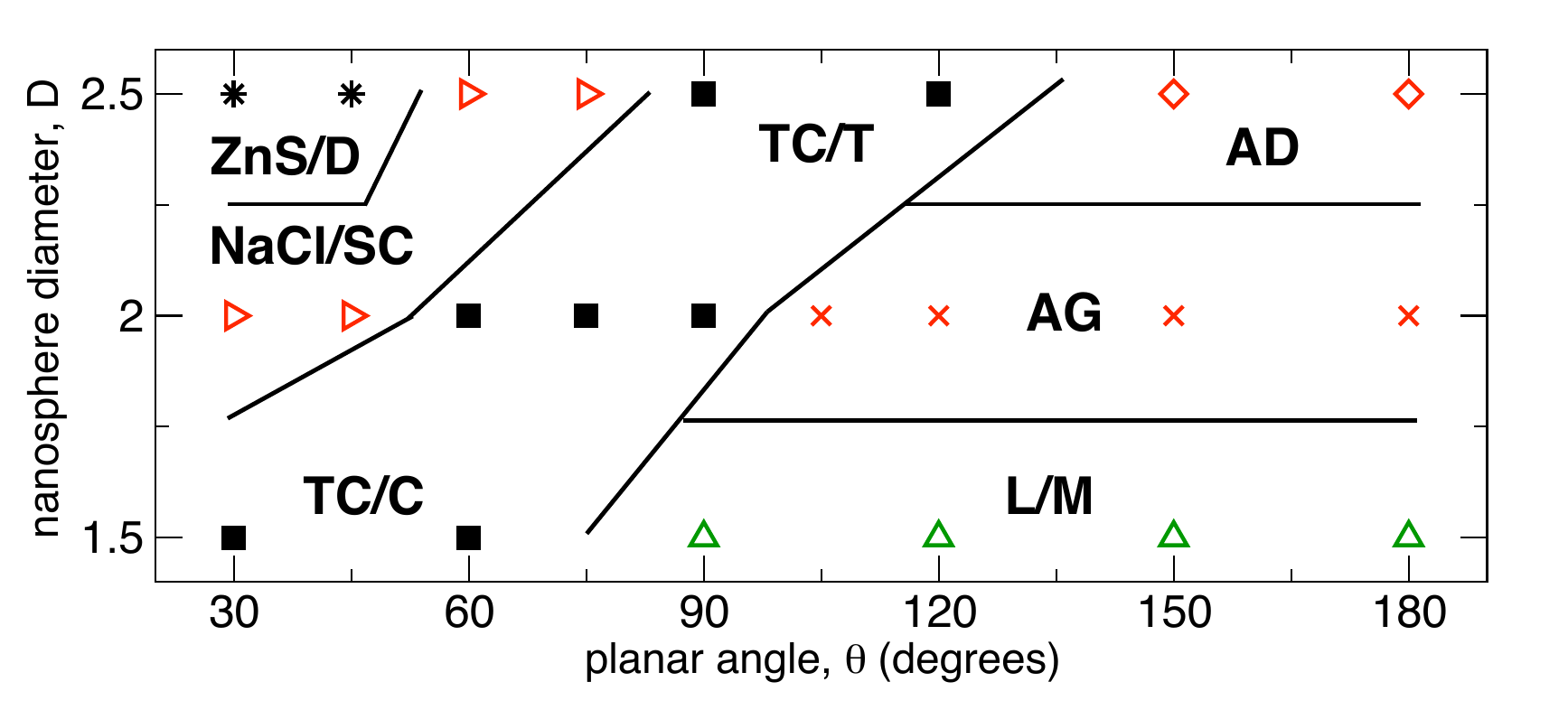} 
\caption{Phase diagram for $D$ verses $\theta$ at 1/T* = 0.8, created from the data presented in figure \ref{DTNS:angle_phasediagram}.  Phase boundaries are approximate and drawn to guide the eye.}
\label{DTNS:angle_diagrams_c}
\end{center}
\end{figure}

The overall phase behavior as a function of nanosphere diameter can be better understood by constructing a plot of $D$ verses $\theta$ for a fixed 1/T*.  In figure \ref{DTNS:angle_diagrams_c} we present the phase behavior for 1/T* = 0.8, which is within the ordered regime for all values of $D$.  We see that spherical micelle phases are only stable in regions where tethers are close together (i.e. small $\theta$) and the nanosphere diameter is ``large.''  Conversely, lamellar phases are stable in regions where tethers are well spaced (i.e. large $\theta$) and nanosphere diameter is ``small.''  This diagram also demonstrates the strong dependence of most of the structures on nanosphere diameter, e.g. a small decrease in nanosphere diameter may drive the system from ZnS/D to NaCl/SC or from AD to AG.  The notable exception is the tetragonally ordered cylinder phase.  Additionally, we can conjecture that many of these phases may be relatively insensitive to non-idealities present in experimental systems.  For example, the L/M phase forms over a very wide range of $\theta$ values and thus may be relatively insensitive to polydispersity in bond angle.  Likewise, the tetragonally ordered cylinder phases (TC/C and TC/T) form for all $D$ values explored, thus they may be less sensitive to polydispersity in nanosphere diameter.  

The progression of phases with respect to decreasing $\theta$ is similar to changing the length of the middle block (B block) in a linear triblock copolymer. In linear triblock copolymers, as the length of the middle block is decreased, we typically see a change from lamellae \cite{mogi1992, dotera1996, matsen1998} to a tricontinuous structure (e.g. the alternating gyroid \cite{matsen1998} or alternating diamond phase \cite{dotera1996, mogidiamon1992, mogi1992}) to alternating tetragonal cylinders \cite{matsen1998, mogi1992} to CsCl structured micelles \cite{matsen1998, mogi1992}.  In both cases, whether changing $\theta$ or the length of the middle block, the net result is that the A and C blocks are brought closer together, constraining the possible tether configurations and resulting in a phase transition.   Additionally, we see that decreasing the size of the nanospheres drives the system to form structures with less interfacial curvature (e.g. lamellae).
 
The lamellae, alternating diamond, alternating gyroid, and alternating tetragonal cylinder phases are all well known in the linear triblock copolymer literature and their formation in the DTNS system is not entirely surprising; in these cases, the DTNS behave very similar to linear triblocks and the geometry of the nanoparticle appears to have little impact on the overall bulk structure. However, linear triblock copolymers have been shown to form CsCl ordered spherical micelles \cite{matsen1998, phan1998} rather than the NaCl or ZnS ordered micelles we find for DTNS. It has been calculated for linear triblock copolymers that CsCl has a lower free energy than NaCl \cite{phan1998}.   Thus, we would expect that DTNS might also form the CsCl structure and the fact it is not present is somewhat surprising.  As we discussed in reference \cite{iacovella2009}, a nanosphere has no conformational degrees of freedom as compared to an equivalent flexible polymer with the same excluded volume; the radius and effective volume (i.e. shape or mass distribution) of a nanosphere is constant, whereas the radius of gyration and effective volume of a flexible polymer may vary based on solvent conditions, temperature, or volume fraction.  As a result, the flexible middle block in a triblock copolymer allows the A and C blocks a large amount of conformational freedom by changing its conformation.  In the DTNS system, we remove many of the degrees of freedom of the middle block by replacing it with a nanosphere that has a fixed volume/geometric contribution and by including a bond angle constraint, $\theta$.  This limits the conformational entropy of the A and C tethers as compared to the triblock system.  As such, these constraints lead to the stabilization of the NaCl structure rather than CsCl.
\begin{figure}[ht]
\begin{center}
\includegraphics[width=2.25in]{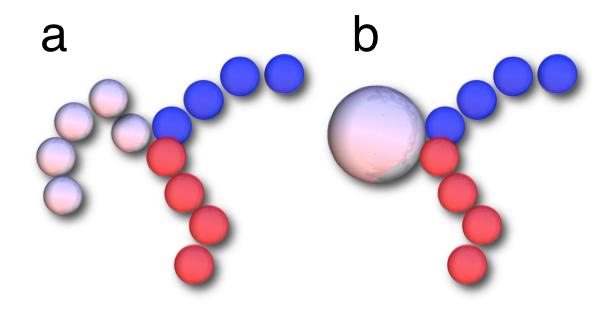} 
\caption{\textbf{(a)} Schematic of a star triblock copolymer. \textbf{(b)} Schematic of DTNS for $\theta$=30$^\circ$.}
\label{DTNS:starblock}
\end{center}
\end{figure}

Similarly, the geometric and architectural constraints of the DTNS are important to the formation of the ZnS/D structure.  In the region where we find the ZnS/D phase, the DTNS building block closely resembles a star triblock copolymer, as sketched in figure \ref{DTNS:starblock}.  In reference \cite{iacovella2009} we showed that the geometry of the nanosphere was crucial to the formation of the ZnS/D structure over the commonly observed [6,6,6] columnar structure.  The geometry of  the nanosphere induces a larger amount of curvature than an equivalent flexible block in a star triblock copolymer, resulting in spherical, rather than cylindrical, morphologies \cite{iacovella2009}.  This conclusion is supported by recent work that showed ZnS ordered spherical micelles for a system composed of a mixture of ABC star triblock copolymers and B homopolymers \cite{dotera2008}.  The addition of the B homopolymer acts to increase the bulk of the B block in the star triblock rather than just the length. This is similar to the impact of the geometry of the nanospheres and highlights the importance that the distribution of volume (i.e. the shape) can have on the overall structure. 

\section{Summary}
\label{conclusion}
Our results show that the structural phase behavior of DTNS is a function of the directionality and strength of the tether-tether interactions and diameter of the nanosphere.  We have shown that DTNS can produce unique structural arrangements of both tethers and nanoparticles that are similar in bulk structure to triblock copolymers.  We have demonstrated a route to form diamond and SC networks of nanoparticles, two structures highly sought for photonic applications\cite{maldovan2002} as well as sheets, cylinders, and tetragonal meshes of nanospheres.  Overall, we have shown that the use of soft-matter tethers with directionality can be used to produce highly ordered periodic structures that would not necessarily be expected of either equivalent flexible polymer systems or pure nanoparticle systems in the absence of tethers.

We thank A.S. Keys for useful discussions, the DOE Grant No. DE-FG02-02ER46000 and the University of Michigan Rackham Predoctoral Fellowship for funding.

\bibliographystyle{rsc}
\bibliography{DTNS}

\providecommand*{\mcitethebibliography}{\thebibliography}
\csname @ifundefined\endcsname{endmcitethebibliography}
{\let\endmcitethebibliography\endthebibliography}{}
\begin{mcitethebibliography}{53}
\providecommand*{\natexlab}[1]{#1}
\providecommand*{\mciteSetBstSublistMode}[1]{}
\providecommand*{\mciteSetBstMaxWidthForm}[2]{}
\providecommand*{\mciteBstWouldAddEndPuncttrue}
  {\def\EndOfBibitem{\unskip.}}
\providecommand*{\mciteBstWouldAddEndPunctfalse}
  {\let\EndOfBibitem\relax}
\providecommand*{\mciteSetBstMidEndSepPunct}[3]{}
\providecommand*{\mciteSetBstSublistLabelBeginEnd}[3]{}
\providecommand*{\EndOfBibitem}{}
\mciteSetBstSublistMode{f}
\mciteSetBstMaxWidthForm{subitem}
{(\emph{\alph{mcitesubitemcount}})}
\mciteSetBstSublistLabelBeginEnd{\mcitemaxwidthsubitemform\space}
{\relax}{\relax}

\bibitem[Hermanson \emph{et~al.}(2001)Hermanson, Lumsdon, Williams, Kaler, and
  Velev]{hermanson2001}
K.~D. Hermanson, S.~O. Lumsdon, J.~P. Williams, E.~W. Kaler and O.~D. Velev,
  \emph{Science}, 2001, \textbf{294}, 1082--1086\relax
\mciteBstWouldAddEndPuncttrue
\mciteSetBstMidEndSepPunct{\mcitedefaultmidpunct}
{\mcitedefaultendpunct}{\mcitedefaultseppunct}\relax
\EndOfBibitem
\bibitem[Crossland \emph{et~al.}(0)Crossland, Kamperman, Nedelcu, Ducati,
  Wiesner, Smilgies, Toombes, Hillmyer, Ludwigs, Steiner, and
  Snaith]{gyroidsolarcell}
E.~J.~W. Crossland, M.~Kamperman, M.~Nedelcu, C.~Ducati, U.~Wiesner, D.~M.
  Smilgies, G.~E.~S. Toombes, M.~A. Hillmyer, S.~Ludwigs, U.~Steiner and H.~J.
  Snaith, \emph{Nano Letters}, 0, \textbf{0}, year\relax
\mciteBstWouldAddEndPuncttrue
\mciteSetBstMidEndSepPunct{\mcitedefaultmidpunct}
{\mcitedefaultendpunct}{\mcitedefaultseppunct}\relax
\EndOfBibitem
\bibitem[Zhang \emph{et~al.}(2006)Zhang, Sun, Friend, Guo, Nau, and
  Giessen]{zhang2006}
X.~P. Zhang, B.~Q. Sun, R.~H. Friend, H.~C. Guo, D.~Nau and H.~Giessen,
  \emph{Nano Letters}, 2006, \textbf{6}, 651--655\relax
\mciteBstWouldAddEndPuncttrue
\mciteSetBstMidEndSepPunct{\mcitedefaultmidpunct}
{\mcitedefaultendpunct}{\mcitedefaultseppunct}\relax
\EndOfBibitem
\bibitem[Shevchenko \emph{et~al.}(2008)Shevchenko, Ringler, Schwemer, Talapin,
  Klar, Rogach, Feldmann, and Alivisatos]{shevchenko2008}
E.~V. Shevchenko, M.~Ringler, A.~Schwemer, D.~V. Talapin, T.~A. Klar, A.~L.
  Rogach, J.~Feldmann and A.~P. Alivisatos, \emph{Journal of the American
  Chemical Society}, 2008, \textbf{130}, 3274\relax
\mciteBstWouldAddEndPuncttrue
\mciteSetBstMidEndSepPunct{\mcitedefaultmidpunct}
{\mcitedefaultendpunct}{\mcitedefaultseppunct}\relax
\EndOfBibitem
\bibitem[Zhang \emph{et~al.}(2007)Zhang, Tang, Kotov, and Glotzer]{zhang2007}
Z.-L. Zhang, Z.~Y. Tang, N.~A. Kotov and S.~C. Glotzer, \emph{Nano Letters},
  2007, \textbf{7}, 1670--1675\relax
\mciteBstWouldAddEndPuncttrue
\mciteSetBstMidEndSepPunct{\mcitedefaultmidpunct}
{\mcitedefaultendpunct}{\mcitedefaultseppunct}\relax
\EndOfBibitem
\bibitem[Tang \emph{et~al.}(2006)Tang, Zhang, Wang, Glotzer, and
  Kotov]{tang2006}
Z.~Y. Tang, Z.-L. Zhang, Y.~Wang, S.~C. Glotzer and N.~A. Kotov,
  \emph{Science}, 2006, \textbf{314}, 274--278\relax
\mciteBstWouldAddEndPuncttrue
\mciteSetBstMidEndSepPunct{\mcitedefaultmidpunct}
{\mcitedefaultendpunct}{\mcitedefaultseppunct}\relax
\EndOfBibitem
\bibitem[Tang \emph{et~al.}(2002)Tang, Kotov, and Giersig]{tang2002}
Z.~Y. Tang, N.~A. Kotov and M.~Giersig, \emph{Science}, 2002, \textbf{297},
  237--240\relax
\mciteBstWouldAddEndPuncttrue
\mciteSetBstMidEndSepPunct{\mcitedefaultmidpunct}
{\mcitedefaultendpunct}{\mcitedefaultseppunct}\relax
\EndOfBibitem
\bibitem[Cho \emph{et~al.}(2005)Cho, Talapin, Gaschler, and
  Murray]{murraydipole2005}
K.~S. Cho, D.~V. Talapin, W.~Gaschler and C.~B. Murray, \emph{Journal of the
  American Chemical Society}, 2005, \textbf{127}, 7140--7147\relax
\mciteBstWouldAddEndPuncttrue
\mciteSetBstMidEndSepPunct{\mcitedefaultmidpunct}
{\mcitedefaultendpunct}{\mcitedefaultseppunct}\relax
\EndOfBibitem
\bibitem[Shevchenko \emph{et~al.}(2006)Shevchenko, Talapin, Kotov, O'Brien, and
  Murray]{murray2006}
E.~V. Shevchenko, D.~V. Talapin, N.~A. Kotov, S.~O'Brien and C.~B. Murray,
  \emph{Nature}, 2006, \textbf{439}, 55--59\relax
\mciteBstWouldAddEndPuncttrue
\mciteSetBstMidEndSepPunct{\mcitedefaultmidpunct}
{\mcitedefaultendpunct}{\mcitedefaultseppunct}\relax
\EndOfBibitem
\bibitem[Knorowski \emph{et~al.}(2008)Knorowski, Anderson, and
  Travesset]{knorowski2008}
C.~D. Knorowski, J.~A. Anderson and A.~Travesset, \emph{JOURNAL OF CHEMICAL
  PHYSICS}, 2008, \textbf{128}, year\relax
\mciteBstWouldAddEndPuncttrue
\mciteSetBstMidEndSepPunct{\mcitedefaultmidpunct}
{\mcitedefaultendpunct}{\mcitedefaultseppunct}\relax
\EndOfBibitem
\bibitem[Zhang \emph{et~al.}(2003)Zhang, Horsch, Lamm, and Glotzer]{zhang2003}
Z.-L. Zhang, M.~Horsch, M.~H. Lamm and S.~C. Glotzer, \emph{Nano Letters},
  2003, \textbf{3}, 1341--1346\relax
\mciteBstWouldAddEndPuncttrue
\mciteSetBstMidEndSepPunct{\mcitedefaultmidpunct}
{\mcitedefaultendpunct}{\mcitedefaultseppunct}\relax
\EndOfBibitem
\bibitem[Horsch \emph{et~al.}(2005)Horsch, Zhang, and Glotzer]{horsch2005}
M.~A. Horsch, Z.~L. Zhang and S.~C. Glotzer, \emph{Physical Review Letters},
  2005, \textbf{95}, 056105\relax
\mciteBstWouldAddEndPuncttrue
\mciteSetBstMidEndSepPunct{\mcitedefaultmidpunct}
{\mcitedefaultendpunct}{\mcitedefaultseppunct}\relax
\EndOfBibitem
\bibitem[Horsch \emph{et~al.}(2006)Horsch, Zhang, and Glotzer]{horsch2006}
M.~A. Horsch, Z.~L. Zhang and S.~C. Glotzer, \emph{Journal of Chemical
  Physics}, 2006, \textbf{125}, year\relax
\mciteBstWouldAddEndPuncttrue
\mciteSetBstMidEndSepPunct{\mcitedefaultmidpunct}
{\mcitedefaultendpunct}{\mcitedefaultseppunct}\relax
\EndOfBibitem
\bibitem[Iacovella \emph{et~al.}(2005)Iacovella, Horsch, Zhang, and
  Glotzer]{iacovella2005}
C.~R. Iacovella, M.~A. Horsch, Z.~Zhang and S.~C. Glotzer, \emph{Langmuir},
  2005, \textbf{21}, 9488--9494\relax
\mciteBstWouldAddEndPuncttrue
\mciteSetBstMidEndSepPunct{\mcitedefaultmidpunct}
{\mcitedefaultendpunct}{\mcitedefaultseppunct}\relax
\EndOfBibitem
\bibitem[Iacovella \emph{et~al.}(2007)Iacovella, Keys, Horsch, and
  Glotzer]{iacovella2007}
C.~R. Iacovella, A.~S. Keys, M.~A. Horsch and S.~C. Glotzer, \emph{Physical
  Review E}, 2007, \textbf{75}, 040801\relax
\mciteBstWouldAddEndPuncttrue
\mciteSetBstMidEndSepPunct{\mcitedefaultmidpunct}
{\mcitedefaultendpunct}{\mcitedefaultseppunct}\relax
\EndOfBibitem
\bibitem[Iacovella \emph{et~al.}(2008)Iacovella, Horsch, and
  Glotzer]{iacovella2008}
C.~R. Iacovella, M.~A. Horsch and S.~C. Glotzer, \emph{Journal of Chemical
  Physics}, 2008, \textbf{129}, 044902\relax
\mciteBstWouldAddEndPuncttrue
\mciteSetBstMidEndSepPunct{\mcitedefaultmidpunct}
{\mcitedefaultendpunct}{\mcitedefaultseppunct}\relax
\EndOfBibitem
\bibitem[Iacovella and Glotzer(2009)]{iacovella2009}
C.~R. Iacovella and S.~C. Glotzer, \emph{Nano Letters}, 2009, \textbf{9},
  1206--1211\relax
\mciteBstWouldAddEndPuncttrue
\mciteSetBstMidEndSepPunct{\mcitedefaultmidpunct}
{\mcitedefaultendpunct}{\mcitedefaultseppunct}\relax
\EndOfBibitem
\bibitem[Jayaraman and Schweizer(2008)]{arthi2008}
A.~Jayaraman and K.~S. Schweizer, \emph{Journal of Chemical Physics}, 2008,
  \textbf{128}, year\relax
\mciteBstWouldAddEndPuncttrue
\mciteSetBstMidEndSepPunct{\mcitedefaultmidpunct}
{\mcitedefaultendpunct}{\mcitedefaultseppunct}\relax
\EndOfBibitem
\bibitem[Jayaraman and Schweizer(2008)]{arthi2008b}
A.~Jayaraman and K.~S. Schweizer, \emph{Langmuir}, 2008, \textbf{24},
  11119--11130\relax
\mciteBstWouldAddEndPuncttrue
\mciteSetBstMidEndSepPunct{\mcitedefaultmidpunct}
{\mcitedefaultendpunct}{\mcitedefaultseppunct}\relax
\EndOfBibitem
\bibitem[Wilson \emph{et~al.}(2009)Wilson, Thomas, Dennison, and
  Masters]{wilson2009}
M.~R. Wilson, A.~B. Thomas, M.~Dennison and A.~J. Masters, \emph{SOFT MATTER},
  2009, \textbf{5}, 363--368\relax
\mciteBstWouldAddEndPuncttrue
\mciteSetBstMidEndSepPunct{\mcitedefaultmidpunct}
{\mcitedefaultendpunct}{\mcitedefaultseppunct}\relax
\EndOfBibitem
\bibitem[DeVries \emph{et~al.}(2007)DeVries, Brunnbauer, Hu, Jackson, Long,
  Neltner, Uzun, Wunsch, and Stellacci]{devries2007}
G.~A. DeVries, M.~Brunnbauer, Y.~Hu, A.~M. Jackson, B.~Long, B.~T. Neltner,
  O.~Uzun, B.~H. Wunsch and F.~Stellacci, \emph{Science}, 2007, \textbf{315},
  358--361\relax
\mciteBstWouldAddEndPuncttrue
\mciteSetBstMidEndSepPunct{\mcitedefaultmidpunct}
{\mcitedefaultendpunct}{\mcitedefaultseppunct}\relax
\EndOfBibitem
\bibitem[DeVries \emph{et~al.}(2008)DeVries, Talley, Carney, and
  Stellacci]{devries2008}
G.~A. DeVries, F.~R. Talley, R.~P. Carney and F.~Stellacci, \emph{Advanced
  Materials}, 2008, \textbf{20}, 4243--4247\relax
\mciteBstWouldAddEndPuncttrue
\mciteSetBstMidEndSepPunct{\mcitedefaultmidpunct}
{\mcitedefaultendpunct}{\mcitedefaultseppunct}\relax
\EndOfBibitem
\bibitem[Nie \emph{et~al.}(2007)Nie, Fava, Kumacheva, Zou, Walker, and
  Rubinstein]{nie2007}
Z.~H. Nie, D.~Fava, E.~Kumacheva, S.~Zou, G.~C. Walker and M.~Rubinstein,
  \emph{Nature Materials}, 2007, \textbf{6}, 609--614\relax
\mciteBstWouldAddEndPuncttrue
\mciteSetBstMidEndSepPunct{\mcitedefaultmidpunct}
{\mcitedefaultendpunct}{\mcitedefaultseppunct}\relax
\EndOfBibitem
\bibitem[Kim \emph{et~al.}(2005)Kim, Pyun, Frechet, Hawker, and
  Frank]{frank2005}
Y.~Kim, J.~Pyun, J.~M.~J. Frechet, C.~J. Hawker and C.~W. Frank,
  \emph{Langmuir}, 2005, \textbf{21}, 10444--10458\relax
\mciteBstWouldAddEndPuncttrue
\mciteSetBstMidEndSepPunct{\mcitedefaultmidpunct}
{\mcitedefaultendpunct}{\mcitedefaultseppunct}\relax
\EndOfBibitem
\bibitem[Song \emph{et~al.}(2003)Song, Dai, Tam, Lee, and Goh]{song2003}
T.~Song, S.~Dai, K.~C. Tam, S.~Y. Lee and S.~H. Goh, \emph{Langmuir}, 2003,
  \textbf{19}, 4798--4803\relax
\mciteBstWouldAddEndPuncttrue
\mciteSetBstMidEndSepPunct{\mcitedefaultmidpunct}
{\mcitedefaultendpunct}{\mcitedefaultseppunct}\relax
\EndOfBibitem
\bibitem[Song \emph{et~al.}(2003)Song, Dai, Tam, Lee, and Goh]{song2003b}
T.~Song, S.~Dai, K.~C. Tam, S.~Y. Lee and S.~H. Goh, \emph{Polymer}, 2003,
  \textbf{44}, 2529--2536\relax
\mciteBstWouldAddEndPuncttrue
\mciteSetBstMidEndSepPunct{\mcitedefaultmidpunct}
{\mcitedefaultendpunct}{\mcitedefaultseppunct}\relax
\EndOfBibitem
\bibitem[Nguyen \emph{et~al.}(2008)Nguyen, Zhang, and Glotzer]{trung2008}
T.~D. Nguyen, Z.~Zhang and S.~C. Glotzer, \emph{Journal of Chemical Physics},
  2008, \textbf{129}, 244903\relax
\mciteBstWouldAddEndPuncttrue
\mciteSetBstMidEndSepPunct{\mcitedefaultmidpunct}
{\mcitedefaultendpunct}{\mcitedefaultseppunct}\relax
\EndOfBibitem
\bibitem[Chan \emph{et~al.}(2005)Chan, Zhang, Lee, Neurock, and
  Glotzer]{chan2005}
E.~R. Chan, X.~Zhang, C.~Y. Lee, M.~Neurock and S.~C. Glotzer,
  \emph{Macromolecules}, 2005, \textbf{38}, 14\relax
\mciteBstWouldAddEndPuncttrue
\mciteSetBstMidEndSepPunct{\mcitedefaultmidpunct}
{\mcitedefaultendpunct}{\mcitedefaultseppunct}\relax
\EndOfBibitem
\bibitem[Chan \emph{et~al.}(2006)Chan, Ho, and Glotzer]{chan2006}
E.~R. Chan, L.~C. Ho and S.~C. Glotzer, \emph{Journal of Chemical Physics},
  2006, \textbf{125}, year\relax
\mciteBstWouldAddEndPuncttrue
\mciteSetBstMidEndSepPunct{\mcitedefaultmidpunct}
{\mcitedefaultendpunct}{\mcitedefaultseppunct}\relax
\EndOfBibitem
\bibitem[Zhang \emph{et~al.}(2005)Zhang, Chan, and Glotzer]{zhangx2005}
X.~Zhang, E.~R. Chan and S.~C. Glotzer, \emph{Journal of Chemical Physics},
  2005, \textbf{123}, year\relax
\mciteBstWouldAddEndPuncttrue
\mciteSetBstMidEndSepPunct{\mcitedefaultmidpunct}
{\mcitedefaultendpunct}{\mcitedefaultseppunct}\relax
\EndOfBibitem
\bibitem[Glotzer \emph{et~al.}(2005)Glotzer, Horsch, Iacovella, Zhang, Chan,
  and Zhang]{glotzer2005}
S.~C. Glotzer, M.~A. Horsch, C.~R. Iacovella, Z.~L. Zhang, E.~R. Chan and
  X.~Zhang, \emph{Current Opinion in Colloid and Interface Science}, 2005,
  \textbf{10}, 287--295\relax
\mciteBstWouldAddEndPuncttrue
\mciteSetBstMidEndSepPunct{\mcitedefaultmidpunct}
{\mcitedefaultendpunct}{\mcitedefaultseppunct}\relax
\EndOfBibitem
\bibitem[Glotzer and Solomon(2007)]{glotzer2007}
S.~C. Glotzer and M.~J. Solomon, \emph{Nature Materials}, 2007, \textbf{6},
  557--562\relax
\mciteBstWouldAddEndPuncttrue
\mciteSetBstMidEndSepPunct{\mcitedefaultmidpunct}
{\mcitedefaultendpunct}{\mcitedefaultseppunct}\relax
\EndOfBibitem
\bibitem[Stellacci()]{stellacci}
F.~Stellacci, Personal Communication\relax
\mciteBstWouldAddEndPuncttrue
\mciteSetBstMidEndSepPunct{\mcitedefaultmidpunct}
{\mcitedefaultendpunct}{\mcitedefaultseppunct}\relax
\EndOfBibitem
\bibitem[Grest and Kremer(1986)]{grest1986}
G.~S. Grest and K.~Kremer, \emph{Physical Review A}, 1986, \textbf{33},
  3628--3631\relax
\mciteBstWouldAddEndPuncttrue
\mciteSetBstMidEndSepPunct{\mcitedefaultmidpunct}
{\mcitedefaultendpunct}{\mcitedefaultseppunct}\relax
\EndOfBibitem
\bibitem[Chandler \emph{et~al.}(1983)Chandler, Weeks, and
  Andersen]{chandler1983}
D.~Chandler, J.~D. Weeks and H.~C. Andersen, \emph{Science}, 1983,
  \textbf{220}, 787--794\relax
\mciteBstWouldAddEndPuncttrue
\mciteSetBstMidEndSepPunct{\mcitedefaultmidpunct}
{\mcitedefaultendpunct}{\mcitedefaultseppunct}\relax
\EndOfBibitem
\bibitem[Crocker and Grier(1996)]{crocker1996}
J.~C. Crocker and D.~G. Grier, \emph{Journal of Colloid and Interface Science},
  1996, \textbf{179}, 298--310\relax
\mciteBstWouldAddEndPuncttrue
\mciteSetBstMidEndSepPunct{\mcitedefaultmidpunct}
{\mcitedefaultendpunct}{\mcitedefaultseppunct}\relax
\EndOfBibitem
\bibitem[Varadan and Solomon(2003)]{varadan2003}
P.~Varadan and M.~J. Solomon, \emph{Langmuir}, 2003, \textbf{19},
  509--512\relax
\mciteBstWouldAddEndPuncttrue
\mciteSetBstMidEndSepPunct{\mcitedefaultmidpunct}
{\mcitedefaultendpunct}{\mcitedefaultseppunct}\relax
\EndOfBibitem
\bibitem[Cyganek and Siebert(2009)]{cyganek2009}
B.~Cyganek and J.~P. Siebert, \emph{An Introduction to 3D Computer Vision
  Techniques and Algorithms}, Wiley, 2009\relax
\mciteBstWouldAddEndPuncttrue
\mciteSetBstMidEndSepPunct{\mcitedefaultmidpunct}
{\mcitedefaultendpunct}{\mcitedefaultseppunct}\relax
\EndOfBibitem
\bibitem[Schultz(2003)]{schultzthesis2003}
A.~Schultz, \emph{Ph.D. thesis}, North Carolina State University, 2003\relax
\mciteBstWouldAddEndPuncttrue
\mciteSetBstMidEndSepPunct{\mcitedefaultmidpunct}
{\mcitedefaultendpunct}{\mcitedefaultseppunct}\relax
\EndOfBibitem
\bibitem[De~Graef and McHenry(2007)]{structureofmaterials2007}
M.~De~Graef and M.~E. McHenry, \emph{Structure of Materials}, Cambridge
  University Press, 2007\relax
\mciteBstWouldAddEndPuncttrue
\mciteSetBstMidEndSepPunct{\mcitedefaultmidpunct}
{\mcitedefaultendpunct}{\mcitedefaultseppunct}\relax
\EndOfBibitem
\bibitem[Martinez-Veracoechea and Escobedo(2005)]{escobedo2005}
F.~J. Martinez-Veracoechea and F.~A. Escobedo, \emph{Macromolecules}, 2005,
  \textbf{38}, 8522--8531\relax
\mciteBstWouldAddEndPuncttrue
\mciteSetBstMidEndSepPunct{\mcitedefaultmidpunct}
{\mcitedefaultendpunct}{\mcitedefaultseppunct}\relax
\EndOfBibitem
\bibitem[Hajduk \emph{et~al.}(1994)Hajduk, Harper, Gruner, Honeker, Kim,
  Thomas, and Fetters]{hajduk1994}
D.~A. Hajduk, P.~E. Harper, S.~M. Gruner, C.~C. Honeker, G.~Kim, E.~L. Thomas
  and L.~J. Fetters, \emph{Macromolecules}, 1994, \textbf{27}, 4063--4075\relax
\mciteBstWouldAddEndPuncttrue
\mciteSetBstMidEndSepPunct{\mcitedefaultmidpunct}
{\mcitedefaultendpunct}{\mcitedefaultseppunct}\relax
\EndOfBibitem
\bibitem[Matsen and Bates(1996)]{matsen1996}
M.~W. Matsen and F.~S. Bates, \emph{Macromolecules}, 1996, \textbf{29},
  7641--7644\relax
\mciteBstWouldAddEndPuncttrue
\mciteSetBstMidEndSepPunct{\mcitedefaultmidpunct}
{\mcitedefaultendpunct}{\mcitedefaultseppunct}\relax
\EndOfBibitem
\bibitem[Martinez-Veracoechea and Escobedo(2007)]{escobedo2007}
F.~J. Martinez-Veracoechea and F.~A. Escobedo, \emph{Macromolecules}, 2007,
  \textbf{40}, 7354--7365\relax
\mciteBstWouldAddEndPuncttrue
\mciteSetBstMidEndSepPunct{\mcitedefaultmidpunct}
{\mcitedefaultendpunct}{\mcitedefaultseppunct}\relax
\EndOfBibitem
\bibitem[nav(2009)]{navysite}
\emph{Center for Computational Materials Science of the United States Naval
  Research Laboratory, Crystal Lattice Structures Web page}, 2009,
  \url{http://cst-www.nrl.navy.mil/lattice/}\relax
\mciteBstWouldAddEndPuncttrue
\mciteSetBstMidEndSepPunct{\mcitedefaultmidpunct}
{\mcitedefaultendpunct}{\mcitedefaultseppunct}\relax
\EndOfBibitem
\bibitem[Mogi \emph{et~al.}(1992)Mogi, Kotsuji, Kaneko, Matsushita, and
  Noda]{mogi1992}
Y.~Mogi, H.~Kotsuji, Y.~Kaneko, Y.~Matsushita and I.~Noda,
  \emph{Macromolecules}, 1992, \textbf{25}, 5408--5411\relax
\mciteBstWouldAddEndPuncttrue
\mciteSetBstMidEndSepPunct{\mcitedefaultmidpunct}
{\mcitedefaultendpunct}{\mcitedefaultseppunct}\relax
\EndOfBibitem
\bibitem[Dotera and Hatano(1996)]{dotera1996}
T.~Dotera and A.~Hatano, \emph{Journal of Chemical Physics}, 1996,
  \textbf{105}, 8413--1427\relax
\mciteBstWouldAddEndPuncttrue
\mciteSetBstMidEndSepPunct{\mcitedefaultmidpunct}
{\mcitedefaultendpunct}{\mcitedefaultseppunct}\relax
\EndOfBibitem
\bibitem[Matsen(1998)]{matsen1998}
M.~W. Matsen, \emph{Journal of Chemical Physics}, 1998, \textbf{108},
  785--796\relax
\mciteBstWouldAddEndPuncttrue
\mciteSetBstMidEndSepPunct{\mcitedefaultmidpunct}
{\mcitedefaultendpunct}{\mcitedefaultseppunct}\relax
\EndOfBibitem
\bibitem[Mogi \emph{et~al.}(1992)Mogi, Mori, Matsushita, and
  Noda]{mogidiamon1992}
Y.~Mogi, K.~Mori, Y.~Matsushita and I.~Noda, \emph{Macromolecules}, 1992,
  \textbf{25}, 5412--5415\relax
\mciteBstWouldAddEndPuncttrue
\mciteSetBstMidEndSepPunct{\mcitedefaultmidpunct}
{\mcitedefaultendpunct}{\mcitedefaultseppunct}\relax
\EndOfBibitem
\bibitem[Phan and Fredrickson(1998)]{phan1998}
S.~Phan and G.~H. Fredrickson, \emph{Macromolecules}, 1998, \textbf{31},
  59--63\relax
\mciteBstWouldAddEndPuncttrue
\mciteSetBstMidEndSepPunct{\mcitedefaultmidpunct}
{\mcitedefaultendpunct}{\mcitedefaultseppunct}\relax
\EndOfBibitem
\bibitem[Hayashida \emph{et~al.}(2008)Hayashida, Takano, Dotera, and
  Matsushita]{dotera2008}
K.~Hayashida, A.~Takano, T.~Dotera and Y.~Matsushita, \emph{Macromolecules},
  2008, \textbf{41}, 6269--6271\relax
\mciteBstWouldAddEndPuncttrue
\mciteSetBstMidEndSepPunct{\mcitedefaultmidpunct}
{\mcitedefaultendpunct}{\mcitedefaultseppunct}\relax
\EndOfBibitem
\bibitem[Maldovan \emph{et~al.}(2002)Maldovan, Urbas, Yufa, Carter, and
  Thomas]{maldovan2002}
M.~Maldovan, A.~M. Urbas, N.~Yufa, W.~C. Carter and E.~L. Thomas,
  \emph{Physical Review B}, 2002, \textbf{65}, \relax
\mciteBstWouldAddEndPuncttrue
\mciteSetBstMidEndSepPunct{\mcitedefaultmidpunct}
{\mcitedefaultendpunct}{\mcitedefaultseppunct}\relax
\EndOfBibitem
\end{mcitethebibliography}

\end{document}